\documentclass[twocolumn,aps,prl,showpacs]{revtex4-1}
\usepackage{graphicx}
\usepackage{epsf}
\usepackage{amsmath}
\usepackage{amssymb}

\DeclareMathAlphabet{\mathpzc}{OT1}{pzc}{m}{it}

\begin{document}

\title{An optimal topological spin pump}

\author{Dganit Meidan}
\affiliation{Dahlem Center for Complex Quantum Systems and Institut
f\"{u}r Theoretische Physik, Freie Universit\"{a}t Berlin, 14195
Berlin, Germany}

\author{Tobias Micklitz~\footnote{The first two authors contributed equally to this work}}
\affiliation{Dahlem Center for Complex Quantum Systems and Institut
f\"{u}r Theoretische Physik, Freie Universit\"{a}t Berlin, 14195
Berlin, Germany}

\author{Piet W.~Brouwer}
\affiliation{Dahlem Center for Complex Quantum Systems and Institut
f\"{u}r Theoretische Physik, Freie Universit\"{a}t Berlin, 14195
Berlin, Germany}

\date{\today}

\begin{abstract}

We study the recently introduced  ${\mathbb Z}_2$  pump consisting of a family of one-dimensional bulk insulators with time reversal restriction on the pumping cycle. We find that the scattering matrices of these pumps are dichotomized by a topological index. We show that the class of pumps characterized by a nontrivial
topological index allows, in contrast to its topologically trivial counterpart, for the noiseless pumping of quantized spin,
even in the absence of  spin conservation.  
This distinction sheds light on the   ${\mathbb Z}_2$   classification of two-dimensional time reversal invariant insulators. 

\end{abstract}
\maketitle

{\it Introduction:---}
The idea to pump spin through a mesoscopic device 
at zero bias by cyclic variation of two 
system parameters is very appealing due to its promise 
of precise and reversible flow control~\cite{pumps}.
%\cite{PSharmaPRL2001,ERMuccioloPRL2002,TAonoPRB2003, MGovernalePRB2003,PSharmaPRL2003,RCitroPRB2006,ASchillerPRB2008}. 
{\it Optimal pumps} 
which operate noiseless, transferring quantized spin in a 
cycle are particularly relevant for potential applications~\cite{OEntinWohlmanPRB2002}.
In this paper, we propose the concept of an optimal topological spin pump,
composed of a bulk insulator. We show that the
ability to pump quantized noiseless spin is a hallmark of the
nontrivial  topological invariant characterizing a quantum spin
Hall system.

The discovery of  the quantum Hall effect introduced an 
interesting new way to classify different states of matter. Unlike more 
familiar phases, a quantum Hall state does not break any 
symmetry and cannot be described by a local order parameter. 
Rather, it differs from a regular two-dimensional insulator by 
topological invariants, known as Chern numbers, which reflect 
the global structure of its ground state wave function~\cite{DJThoulessPRL1982,XGWenADP1995}. 
This topological classification of states of matter found recently
an exciting extension to two-dimensional time-reversal invariant bulk 
insulators~\cite{CLKanePRL2005b,LFuPRB2006,topo_ins}. 
In accordance with the Chern numbers used to classify quantum Hall 
systems~\cite{CLKanePRL2005b}, these systems can be characterized 
by a ${\mathbb Z}_2 $ topological index, 
based on the properties of their bulk ground state.  
The group of insulators described by a nontrivial ${\mathbb Z}_2 $ 
index are known as quantum spin Hall systems. 

In analogy to the quantum Hall state, it is possible to gain insight into this 
topological classification by studying a pump formed by placing the two-dimensional system on a cylinder threaded by a magnetic flux 
\cite{DJThoulessPRB1983,QNiuJPA1984}. The resulting system can be 
mapped onto a set of one-dimensional time dependent Hamiltonians by 
identifying the magnetic flux with time. This mapping defines a 
${\mathbb Z}_2$ pump~\cite{LFuPRB2006}. 
Nonetheless, in the absence of spin rotation invariance, the physical meaning 
of the pumped ${\mathbb Z}_2 $ charge remains elusive. In this letter we study 
the scattering matrix of the open one-dimensional pump constructed in this way.  
We find that mapping the two-dimensional insulating system onto a 
one-dimensional pump establishes a ${\mathbb Z}_2 $ classification of the scattering 
matrix. The resulting one-dimensional pumps are dichotomized. The family 
of scattering matrices belonging to the topological nontrivial class allows, 
in contrast to its topological trivial counterpart, for the noiseless pumping of 
a quantized spin, even in the absence of spin conservation. 
We illustrate these ideas by two examples.

{\it ${\mathbb Z}_2$-index:---}We consider a family of one-dimensional 
Hamiltonians of non-interacting electrons 
with a bulk energy gap, that depend continuously on a cyclic 
pumping parameter $ t$, $H(t+T)=H(t)$, and satisfy:
\begin{align}
\label{time reversal transformations}
H(-t)&= \sigma_y H^T(t)\sigma_y,
\end{align} 
where $ \sigma_i$ are the Pauli matrices. 
These systems can be viewed as a mapping of the set of two-dimensional 
time-reversal invariant insulators placed on a cylinder, where $t $ 
corresponds to a magnetic flux threading the cylinder.  
Indeed, Eq.~(\ref{time reversal transformations}) implies time-reversal invariance 
of the corresponding two-dimensional system, as is evident upon identification 
$(k_x,t)\to(k_x,k_y)$.
In the context of pumping, Eq. (\ref{time reversal transformations}) 
ensures the existence of two time-reversal invariant moments (TRIM) 
$t_1=0$ and $t_2=T/2$, 
at which $H(t_i)=\sigma_y H^T(t_i)\sigma_y$ where $i=1,2$.
Upon coupling the one-dimensional system to two single channel leads, 
the open system is described in terms of 
a  time-dependent $4\times4$ unitary scattering 
matrix. 
Provided the system exceeds the attenuation length associated with the bulk 
energy gap, the transmission vanishes, 
and the scattering matrix is block diagonal. Each block, $\hat{r}_\alpha $, is a unitary $ 2\times2$ reflection matrix in spin space, where the index $\alpha=L,R$ refers to the left and the right lead, respectively.  
The average spin injected into lead $\alpha$ during a cycle 
can be expressed in terms of the spin current, 
$\vec{s}_\alpha ={\rm Im}\, \textrm{tr}
\left([dr_\alpha/dt] \vec{\sigma}r_\alpha^\dag\right)$~\cite{PWBrouwerPRB1998},
see \eqref{S-cycle} below. 
It can be readily verified that $\vec{s}_\alpha$ is invariant
under gauge-transformations $\hat{r}_\alpha \to e^{i\varphi_\alpha(t)}\hat{r}_\alpha$. 
This allows us to restrict our analysis to the $U(2)/U(1)\simeq SU(2)$ particle-hole symmetric 
 part of $\hat{r}_\alpha$ which we denote by $ \tilde{r}_\alpha$.

The pumping cycle defined in Eq. (\ref{time reversal transformations}) provides 
a mapping from the periodic time, corresponding to a one-dimensional loop, $S^1$, to the
$SU(2) $ space of particle-hole symmetric reflection matrices, $\tilde{r}_\alpha $.  
Since the  $SU(2)$ group is topological equivalent
to the three-sphere, $S^3 $, one may conclude that a general pumping cycle 
does not distinguish between different topological classes, since all closed 
contours on $S^3$ can be contracted to a single point. 
(The first homotopy class of $S^3 $ is zero, $\Pi_1(S^3)=0$).
However, Eq.(\ref{time reversal transformations}) restricts the mapping by 
associating  $S(T/2+t)=\sigma_yS^T(T/2-t)\sigma_y$. In particular, at the
two TRIM, the $SU(2) $ part of the reflection matrix is given by 
$\tilde{r}_\alpha(t_i)=\pm\openone$. 
This restricted mapping has two topological distinct 
classes, which may be categorized by
\begin{align}
\label{Z2Index}
\tilde{r}_\alpha(0)\tilde{r}_\alpha(T/2)= (-1)^{\nu}\openone.
\end{align} 
Any loop  $\tilde{r}_\alpha(t)$ on the three sphere characterized by $\nu=0 $ 
can be contracted onto a single point, while paths with $\nu=1 $ are fixed by 
two 
distinct points at TRIM and cannot be contracted. 
Following \eqref{time reversal transformations}, any additional points at which $\tilde{r}_\alpha=-\openone $ away from the TRIM occur in pairs.
We may, therefore, equivalently define $\nu$ as the parity of $\tilde{r}_\alpha = -\openone $ moments
traversed in a cycle.  Away from the 
TRIM, however, these points are not protected by symmetry and
can be removed by a perturbation larger than the level broadening introduced by coupling to leads, see Eq.~\eqref{phi}. Hence, in the weak coupling limit, we can disregard any such accidental points.

{\it Topological spin pump:---}  
The ${\mathbb Z}_2 $ classification of the scattering matrix has a direct effect on 
the spin pumped during a cycle. We show below that the family of one-dimensional 
pumps belonging to the topological nontrivial class can, in contrast to its 
topologically trivial counterpart, operate as optimal pumps. To this end consider the effective Hamiltonian, $H=H_L+H_R$, 
for the left and right edge states 
of the bulk insulator 
\begin{align}
\label{heff}
H_\alpha(t) = \mu_\alpha(t) + \vec{h}_\alpha(t)\cdot \vec{\sigma}, \qquad \alpha=L,R.
\end{align} Here we have employed the fact that left and right edge states 
decouple and the Hamiltonian is a 
sum of two independent $2\times 2 $ hermitian 
matrices in spin space, which can be parameterized by a unit matrix and the $3 $ Pauli matrices, 
$\vec{\sigma} $. 
For transparency of the arguments 
we first consider $\mu_\alpha=0$, where 
the spectrum of \eqref{heff}  is 
particle-hole symmetric, with eigenvalues $\pm |\vec{h}_\alpha(t)|\equiv\pm h_\alpha$ 
that vary during the course of a pumping cycle. 
Using the general expression for the scattering matrix $S = 1 + 2i\pi W^\dagger \left( H - i\pi WW^\dagger \right)^{-1} W$, where $W$ is a matrix that describes the coupling between the
leads and the insulator~\cite{Mahaux1969}, and assuming that the leads couple
equally well to up and down spins, so that $W_{\alpha} = w_{\alpha}
\openone$, the reflection matrix of each block is given by
\begin{align}\label{reflection_mtx}
\hat{r}(t)= \tilde{r}(t) &=e^{i\phi(t) \vec{e}_\phi(t)\cdot \vec{\sigma}},
\end{align} 
where we have dropped the lead index $\alpha $ for brevity. 
Here 
\begin{align}
\label{phi}
\cos \phi(t) = {h(t)^2-\Gamma^2 \over h(t)^2 + \Gamma^2},
\qquad 
\sin \phi(t) = {2 \Gamma h(t) \over h(t)^2 + \Gamma^2} 
\end{align} where $\Gamma
= \pi |w|^2$ is the level broadening due to coupling to the leads  
and we introduced the unit vector, 
\begin{align}\label{e_vector}
\vec{e}_\phi(t)=\vec{h}(t)/h(t),
\end{align} which defines a time dependent rotation axis.
We note that while the angle $\phi$ depends on 
the level broadening, the vector $\vec{e}_\phi $ is exclusively determined 
by the effective Hamiltonian. Eq. (\ref{phi}) shows that whenever the edge 
state crosses the Fermi 
level $h(t_i)=0$,  the angle $ \phi= \pi$, and the reflection matrix is 
resonant $\hat{r}=-\openone$.

Following Ref.~\cite{PWBrouwerPRB1998} we express the 
spin injected into lead $\alpha$ 
during a  cycle in terms of the reflection matrix
\begin{align}
\label{S-cycle}
\vec{S}_\alpha
&= {\hbar \over 2\pi} \oint dt\, \vec{s}_\alpha, \qquad
\vec{s}_\alpha(t) = \text{Im }\text{tr} \left( {d \hat{r}_\alpha\over dt}
\vec{\sigma} \hat{r}_\alpha^\dagger \right).
\end{align} 
In general, the rotation axis $\vec{e}_\phi$  varies during the cycling process. As a result, one cannot identify a time independent axis along which spin pumped during a cycle is quantized. The situation, however, changes when the coupling to the leads is weak. We consider the limit $1/T\ll \Gamma\ll E_\Delta $ where the level broadening 
$\Gamma $ is small compared to the gap, yet large compared to the pumping 
rate, in order to allow for the pump to relax between cycles. In this limit the angle, 
\eqref{phi}, remains close to zero, 
$\phi(t)={\cal O}\left(\Gamma/ E_\Delta \right)$, and changes rapidly to $\pi$ 
whenever a gapless edge state appears in the course of a cycle. The time 
duration of this transition can be estimated as
$\delta t \sim \Gamma/(dh(t_i)/dt)\sim (\Gamma /E_\Delta) T\ll T$. In
the weak coupling limit, $\delta t$ can be made arbitrarily
short in comparison to the time scale for variations of $\vec e_{\phi}$,
so that $\vec e_{\phi}(t)$ may be approximated by its value at the center
of the resonance.
The class of pumps with $\nu=1$ cross a single resonance 
at the TRIM. Hence, in the weak coupling limit 
the reflection matrix of a topologically nontrivial pump describes  
a rotation around a fixed axis,
\begin{eqnarray}\label{approximate_reflection_mtx}
\hat{r}(t)&= e^{i\phi(t) \vec{e}_\phi(t_i)\cdot \vec{\sigma}}\left(1+ {\cal O} 
\left(\frac{\Gamma}{E_\Delta}\right) \right)
\end{eqnarray} where $t_i$ is the TRIM at which the resonance occurs.
As a result, the spin injected into lead $\alpha$ by the class of 
topologically nontrivial pumps is quantized  
\begin{align}
\vec{S} 
= \hbar \vec{e}_\phi(t_i)(1+ {\cal O} (\Gamma/E_\Delta))
\end{align}
where the quantization axis is determined by 
microscopic details of the system. 
Conversely, the class of pumps with $\nu=0$ either remains insulating 
during the entire cycle, or traverses two resonances at the TRIM. In the weak 
coupling limit, the former group can be approximated by a constant reflection 
matrix $r(t)\approx \openone $, and thus does not pump spin. The second subgroup 
crosses two resonances during each pumping cycle. Each such resonance is 
associated with the value of the vector $\vec{e}_\phi(t_i) $. The two vectors 
$\vec{e}_\phi(0)$ and $\vec{e}_\phi(T/2)$, however,
need not be aligned. As a result, one cannot identify a \textit{time independent} 
spin direction that would lead to a quantized  
spin pumped trough an insulator with $\nu=0$.

These observations have a direct implication on the spin noise. Following 
Ref.~\cite{AAndreevPRL2000,YMakhlinPRL2001}, the variance of the spin pumped during a cycle, in 
a direction $\vec{e}_q $, is determined by the time dependence of the vector $
\vec{n}_{\vec{e}_q}\cdot \vec{\sigma} = r(t)^\dag\sigma_{\vec{e}_q} r(t)$
and vanishes for a constant $ \vec{n}_{\vec{e}_q}$~\cite{YMakhlinPRL2001}. In the weak 
coupling limit, the reflection matrix of a topologically nontrivial pump, 
Eq.~(\ref{approximate_reflection_mtx}), describes  a rotation around a fixed axis, 
$\vec{e}_\phi(t_i) $. As a result, the vector $ \vec{n}_{\vec{e}_q}(t)$ with  
$\vec{e}_q= \vec{e}_\phi(t_i)$ remains constant during the course of a pumping cycle. 
It follows that the topological nontrivial spin pump allows for the noiseless 
pumping of quantized spin. Conversely, a trivial insulator that crosses two resonances 
during a cycle cannot be parameterized by a time independent vector. 
Consequently, the trivial pump inevitably operates with generation of finite noise.

The above arguments generalize to finite chemical potential $\mu_\alpha$. Once particle-hole symmetry is broken, the energy levels cross the chemical potential at two different moments, $h(t_\pm)=\pm\mu$. From Eq.~\eqref{time reversal transformations} it follows that these occur symmetrically around the TRIM, $t_\pm=t_i\pm \delta t$. By diagonalizing the Hamiltonian at the crossing point, one can show that the width of each transition is determined by $\Gamma $, and in the weak coupling limit, is associated with a fixed vector $\vec{e}_\phi(t_\pm) $. We note that due to Eq.~\eqref{time reversal transformations}, $\vec{e}_\phi(t_+) $ and $\vec{e}_\phi(t_-) $ are co-linear. Hence, $\hat{r}(t) $ describes a rotation around a fixed axis, and 
the spin pumped through a topologically nontrivial pump is noiseless and quantized in a direction which also depends on  $\mu$.

{\it Two examples:---}To illustrate these ideas we next consider two examples which 
demonstrate the difference between the two topological classes of pumps. To model a topologically nontrivial pump we consider the particle-hole symmetric Hamiltonian $  H =H_0+V_b+ V_{\rm st}+V_{\textrm{so}}$ ~\cite{RShinduJPSJ2005,LFuPRB2006}  with
\begin{eqnarray}\label{hamiltonian_of_TI}
 \nonumber
  H_0 &=& \tau_0\sum_{i,\alpha}\left(c^\dag_{i,\alpha}c_{i+1,\alpha}
  +c^\dag_{i+1,\alpha}c_{i,\alpha} \right)\\
 \nonumber
  V_b &=& b(t) \sum_{i,\alpha,\beta}(-1)^i\sigma_{\alpha,\beta}^z c^\dag_{i,\alpha}c_{i,\beta} \\
 \nonumber
  V_{\rm st} &=& \tau_{\rm st}(t) \sum_{i,\alpha}(-1)^i\left(c^\dag_{i,\alpha}c_{i+1,\alpha}
  + c^\dag_{i+1,\alpha}c_{i,\alpha}\right)\\
  V_{\textrm{so}}&=&\sum_{i,\alpha,\beta}i\vec{e}_{\textrm{so}}\vec{\sigma}_{\alpha,\beta}
  \left(c^\dag_{i,\alpha}c_{i+1,\beta}-c^\dag_{i+1,\alpha}c_{i,\beta}\right).
\end{eqnarray}
Here sums run over $N$ sites and spin indices, 
and 
$\tau_{\rm st}(t)=\tau_{\rm}^0\cos(2\pi t/T)$, 
$b(t)=b^0\sin(2\pi t/T)$. To simplify the illustration we choose 
$\vec{e}_{\textrm{so}}=e_{\textrm{so}} \hat{x}$,
such that $\hat{r}(t)$ depends only on two Pauli-matrices and can be 
represented as a point on the two-sphere. Finally, first and last sites 
are coupled to the leads via spin-independent hopping elements 
$W_\alpha =w_\alpha \openone$, while electrons inside the leads are 
described by a tight-binding model, $H_0 $ in \eqref{hamiltonian_of_TI}.

The transmission and reflection coefficients, $\hat{t}_{RL}$ and $\hat{r}_{L}$,
are found from the transfer matrix
${\cal T}_{RL}$ by solving:
\begin{eqnarray}\label{scattering_mtx_coeff}
 \hat{t}_{RL}\left(\!\!\!
    \begin{array}{c}
      e^{ik_Ra (N+2)} \\
     e^{ik_Ra (N+1)}  \\
    \end{array}
  \!\!\!\right)
   &=& {\cal T}_{RL} \left(\!\!
    \begin{array}{c}
      1+ \hat{r}_{L} \\
      e^{-ik_La}+e^{ik_La} \hat{r}_{L}  \\
    \end{array}
  \!\!\right)\!\!
\end{eqnarray}
where $k_{R,L}\approx \pm \pi/(2a) $ are the Fermi wave vectors in 
right and left lead at half filling, and the transfer 
matrix relates the states 
$\begin{pmatrix} \psi_{N+2}, \psi_{N+1}\end{pmatrix}^T
={\cal T}_{RL}\begin{pmatrix} \psi_{0}, \psi_{-1}\end{pmatrix}^T$, and is derived from Eq. \eqref{hamiltonian_of_TI}.
The resulting  $\hat{t}_{RL} $ is exponentially suppressed over the length associated with the gap. To leading orders in  $\delta\tau=\tau_{\rm st}/\tau_0$, 
$\delta b=b/\tau_0$, and $\delta e= e_{\textrm{so}}/\tau_0$  the reflection matrix is given by Eq.~\eqref{reflection_mtx} with:
\begin{align}
\label{param-nu1}
%\nonumber
\!\!\! \vec{e}_\phi= \!\!
\left(0,\frac{\beta_1}{\sqrt{\alpha_1^2+\beta_1^2}},
\frac{-\alpha_1}{\sqrt{\alpha_1^2+\beta_1^2}} \right)^{\!\!T},
\,\,h={b\over \sqrt{\alpha_1^2+\beta_1^2}}.
\end{align}
Here we introduced 
$\alpha_1=\left(\textrm{Re}\Delta_1+2\delta \tau  \right)$ and 
$\beta_1=\left(\textrm{Im}\Delta_1 +2\delta e \right)$, with 
$\Delta_1 =  \sqrt{\delta b^2+4(\delta \tau+i\delta e)^2}$.

\begin{figure}[h]
\begin{center}
\includegraphics[width=0.45\textwidth]{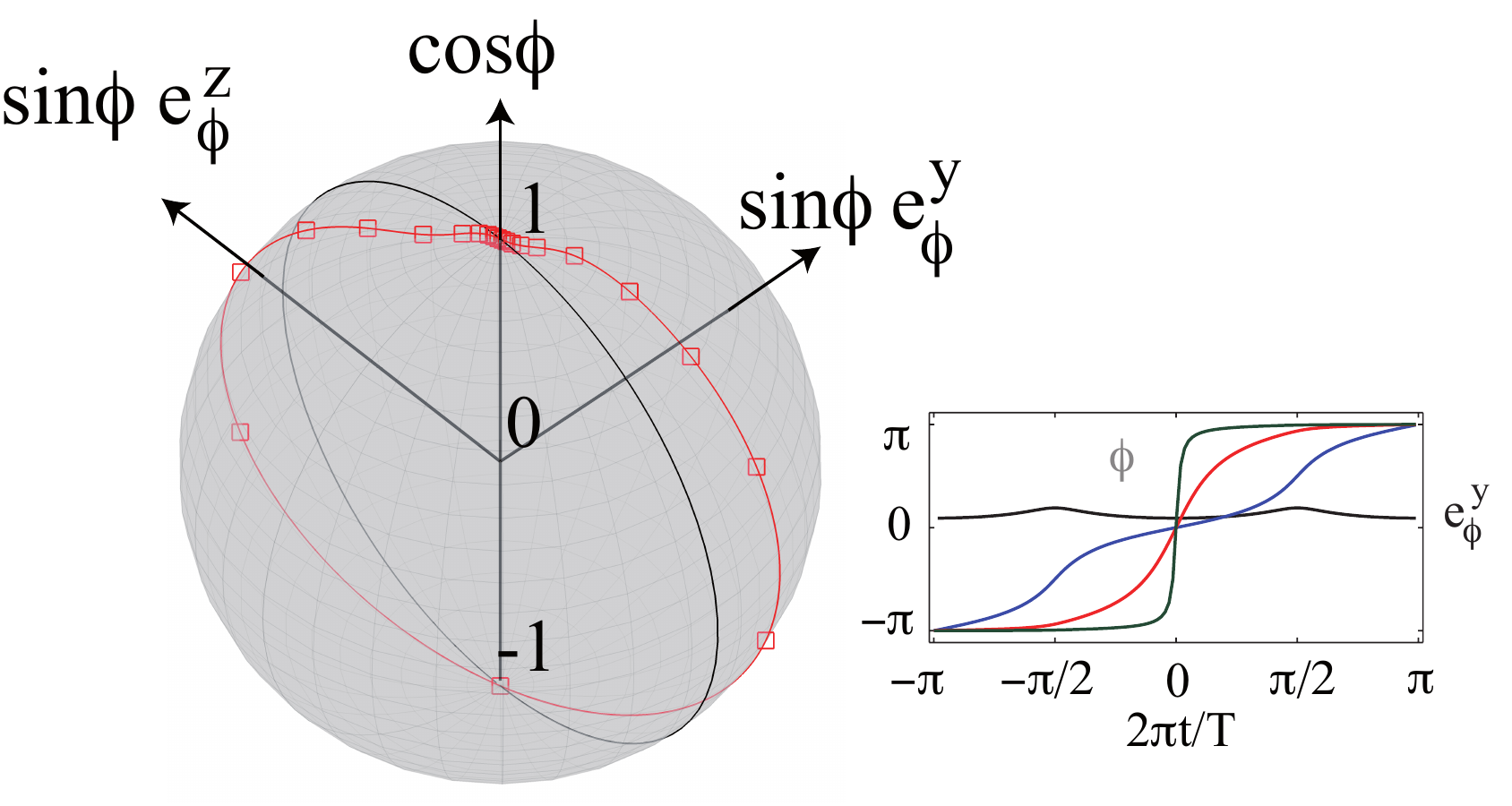}
\caption[0.5\textwidth]{Time evolution of $\hat{r}(t)$ of model ~\eqref{hamiltonian_of_TI}, in the absence of spin orbit (black curve), and for $\delta e/\delta \tau=0.3$ and  $\Gamma/\tau_0 = 0.1$ (red curve). Squares indicate equally spaced time intervals, and illustrate the time duration of the transition to $\hat{r}=-\openone$. The loops cross points $\pm\openone $ at the TRIM $0,T/2 $, respectively, corresponding to a nontrivial ${\mathbb Z}_2$ index $\nu=1$. Inset shows the time dependence  of the $y$-component of $\vec{e}_\phi(t)$ (black curve) and 
of $\phi(t)$ for $\delta e/\delta \tau=0.3 $ and $\Gamma/\tau_0 =1,0.1,0.01 $ (blue red and green curves, respectively). At $\Gamma/\tau_0\rightarrow 0 $, the transition becomes sharper and $\hat{r}(t) $ describes a rotation around a fixed axis, $\vec{e_\phi(T/2)}$, implying that the pump is optimal. }
\label{nu1}
\end{center}
\end{figure}

Fig.~\ref{nu1} shows the time evolution of the reflection matrix, $\hat{r}(t)=\tilde{r}(t)$, of model~\eqref{hamiltonian_of_TI} %during a pumping cycle, 
with and without spin orbit coupling. The inset shows the time dependence of $\phi(t) $ and the $y $ component of the vector $\vec{e}_\phi(t) $, at finite spin orbit coupling and for different coupling strengths. At the TRIM, $t=0,T/2 $, the phase takes the value $\phi=0 $ and $\phi = \pi $, respectively.  Consequently, $\hat{r}(t)$ belongs to the class of pumps with nontrivial 
${\mathbb Z}_2$ index $\nu=1$, see Eq. \eqref{Z2Index}. The loops that represent the time evolution of $\hat{r}(t)$ are, therefore, fixed by the values at the two TRIM, and cannot be 
contracted to a single point by a continuous deformation of 
the microscopic Hamiltonian that  
preserves condition \eqref{time reversal transformations}. In the absence of spin orbit coupling,  $\hat{r}(t)$ 
follows a geodesic, corresponding to the rotation around a fixed axis (black curve). At finite spin orbit coupling, the reflection matrices trace out curved loops, indicating the absence of a fixed rotation axis. At moderately weak coupling,
$\Gamma/\tau_0=0.1$  the reflection matrix follows 
a geodesic during most of the cycle (red curve), whose tangent is determined by the spin orbit vector $\vec{e}_\textrm{so} $. 
In the limit $\Gamma/\tau_0\to 0$ the loops converge
to a geodesic, implying that the pump works optimally. 

To model a topologically trivial pump we consider the Hamiltonian \eqref{hamiltonian_of_TI} with a double spatial periodicity of the time dependent parameters 
\begin{eqnarray}\label{hamiltonian_of_I}
\nonumber
  V_b &=& b(t)\sum_{n,\alpha,\beta}(-1)^n\sigma_{\alpha,\beta}^z (c^\dag_{2 n,\alpha}
  c_{2 n,\beta} +c^\dag_{2 n+1,\alpha}c_{2 n+1,\beta} )\\
  V_{\rm st} &=& \tau_{\rm st}(t) \sum_{n,\alpha}(-1)^n(c^\dag_{2 n+1}c_{2n+2}
  +c^\dag_{2n+2}c_{2n+1}).
 \end{eqnarray} By solving Eq. (\ref{scattering_mtx_coeff}) to leading order in $\delta \tau,\delta b,\delta e$, 
 we find that $\hat r$ is given by
Eq. (\ref{reflection_mtx}) with
\begin{align}
\label{r-nu0}
%\nonumber
\!\!\!\!\vec{e}_\phi\!\! = \!\! \left(\! 0, \! \frac{\delta \tau \beta_2 +2\delta e
\alpha_2 }{\gamma_2\sqrt{\alpha_2^2+\beta_2^2}}, \!\frac{2\delta e \beta_2-\delta \tau
\alpha_2   }{\gamma_2
\sqrt{\alpha_2^2+\beta_2^2}}\right)^{\!\!\!\!T}\!\!,
h\!=\!\!{2b \gamma_2\over \sqrt{\alpha_2^2+\beta_2^2}}.
\!\!\!
\end{align}
Here $\alpha_2 = \textrm{Re}\Delta_2- (\delta b^2-\delta \tau^2)$, 
$\beta_2 = \textrm{Im}\Delta_2 +4\delta e$,  $\Delta_2 = \sqrt{(\delta b^2-\delta \tau^2-4 i \delta e)^2
+(2\delta b\delta \tau-4 i \delta e\delta b)}$ and
$\gamma_2 =\sqrt{ \delta \tau^2+4 \delta e^2}$.

\begin{figure}[h]
\begin{center}
\includegraphics[width=0.45\textwidth]{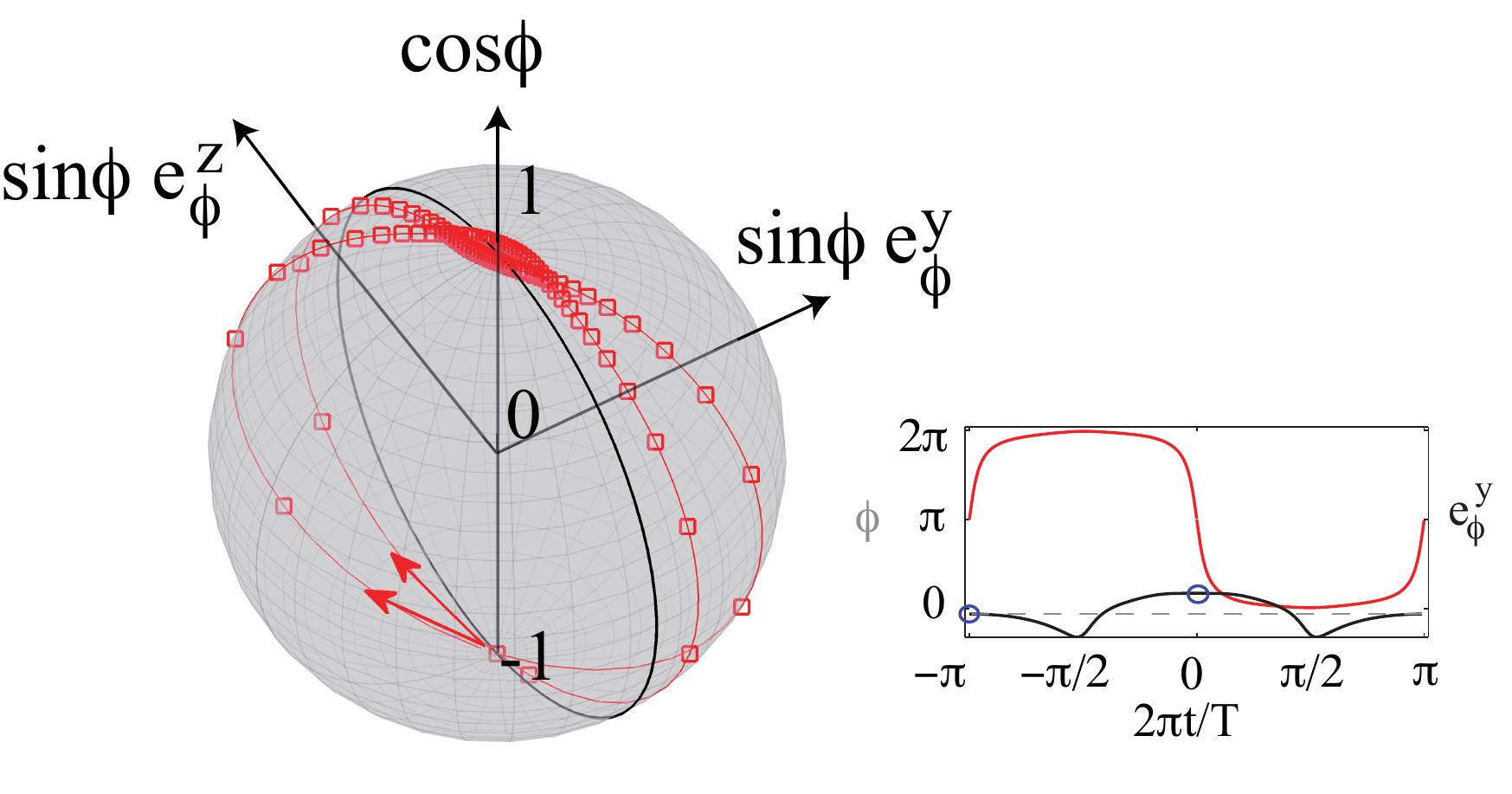}
\caption[0.5\textwidth]{Time evolution of the $\hat{r}(t) $ of model ~\eqref{hamiltonian_of_I}, with ($\delta e/\delta \tau=0.1 $, red curve) and without (black curve) spin orbit coupling, for $\Gamma/\tau_0=0.1 $. The loops cross two resonances during a cycle and thus belong to the trivial class of pumps. At finite spin orbit coupling,  $\hat{r}(t) $ follows two different geodesics, with non-parallel tangents at the TRIM, indicated by the red arrows. 
The inset shows the time dependence of the $y$-component of  $\vec{e}_\phi(t)$ (black curve) and 
of  $\phi(t)$  for $\delta e/\delta \tau=0.1$ and 
$\Gamma/ \tau_0 = 0.1$. The different values of  $\vec{e}_\phi(t_i) $ at the resonances, $ \phi=\pi$,  are marked by the blue circles. }
\label{nu0}
\end{center}
\end{figure}

Fig.~\ref{nu0} shows the time evolution of the reflection matrix, $\hat{r}(t)$, for model~\eqref{hamiltonian_of_I}, with (red curve) and without (black curve) spin orbit coupling. 
The loops cross two resonances at the TRIM, corresponding to the topologically trivial class of pumps, $\nu=0$, Eq. \eqref{Z2Index}. In contrast to the nontrivial class of pumps, these loops are fixed by a single point. As a result, they can be contracted to this single point by a continuous deformation 
of the starting Hamiltonian without violating condition \eqref{time reversal transformations}. This is manifested at finite spin orbit coupling, where the loop avoids the north pole, $\hat{r}=\openone $, while remaining fixed to the south pole, $\hat{r}=-\openone $, as illustrated in the red curve.  
The inset shows the time dependence of the $y$-component of the rotation axis $\vec{e}_\phi(t)$ and 
the angle $\phi(t)$ for finite spin orbit coupling. The angle crosses two resonances, $ \phi=\pi$, during a pumping cycle, which correspond to different values of the vector $\vec{e}_\phi(t_i) $ (see blue circles in the inset).   
The resulting
reflection matrix
traces two loops, which in the limit $\Gamma/\tau_0\to 0$ converge  
to two different geodesics, with non-parallel tangents at the TRIM, indicated by the arrows, see red curve in main figure.  
Hence, in the presence of a finite spin orbit coupling, the spin pumped during a cycle is not quantized, and the pump operates with the generation of finite noise.

{\it Conclusions:---} We have studied the class of one-dimensional pumps
with a time reversal restriction on the pumping cycle, (\ref{time reversal transformations}).  These
systems can be viewed as a mapping of two-dimensional
time-reversal invariant bulk insulators placed on a cylinder, where $t$
corresponds to a magnetic flux threading the cylinder. 
We found that the scattering matrices of the pumps
are dichotomized by the
${\mathbb Z}_2$ topological index.
We have shown that the class of pumps characterized by a nontrivial
topological index allows, in contrast with its topologically trivial counterpart,
for the noiseless pumping of quantized spin, even if spin is not
conserved.  This observation sheds light on the topological classification of two-dimensional time reversal invariant insulators. 

We gratefully acknowledge discussions with F.~von Oppen and
B.~B\'eri. This work is supported by the Alexander von Humboldt Foundation.

\end{document}